# Physics-Informed Neural Network Method for Solving One-Dimensional Advection Equation Using PyTorch.

Shashank Reddy Vadyala; Sai Nethra Betgeri.
Department of Computational Analysis and Modeling, Louisiana Tech University, Ruston, LA United States.

**Abstract:**

Numerical solutions to the equation for advection are determined using different finite-difference approximations and physics-informed neural networks (PINNs) under conditions that allow an analytical solution. Their accuracy is examined by comparing them to the analytical solution. We used a machine learning framework like PyTorch to implement PINNs. PINNs approach allows training neural networks while respecting the PDEs as a strong constraint in the optimization as apposed to making them part of the loss function. In standard small-scale circulation simulations, it is shown that the conventional approach incorporates a pseudo diffusive effect that is almost as large as the effect of the turbulent diffusion model; hence the numerical solution is rendered inconsistent with the PDEs. This oscillation causes inaccuracy and computational uncertainty. Of all the schemes tested, only the PINNs approximation accurately predicted the outcome. We assume that the PINNs approach can transform the physics simulation area by allowing real-time physics simulation and geometry optimization without costly and time-consuming simulations on large supercomputers.

## 1. Introductions

Partially differential equations (PDEs), whose states exist in infinite-dimensional spaces, are often used to model physical science and engineering processes [1]. Due to the lack of computational methods in most situations, finite-dimensional approximations are used instead, which are based on conventional numerical techniques that have been developed and improved over the years. Traditional numerical solvers, on the other hand, frequently necessitate considerable computational effort, particularly for complex systems with multiscale/multiphysics features, and may not be feasible in real-time or many-query applications, such as optimization, inverse problem, and uncertainty quantification (UQ), which necessitate many repeated simulations[2]. Solving PDE structures for the best possible combination between precision and reliability is always a problem [3]. Some data-driven approaches to solve PDEs with deep neural networks (DNNs) exist nowadays. The benefits of using DNNs to approximate PDE solutions[4]. DNNs can Identify nonlinear interactions, which are confirmed mathematically by universal approximation theorems.

Forward assessments of trained DNNs are quick, which is ideal for real-time or multiple-query applications. DNN models are analytically differentiable; derivative information can be easily derived for optimization and control problems using automatic differentiation[5]. Recently, researchers have attempted to use DNNs in numerous sectors, including education[6], media[7], power[8], and healthcare[9]. There are several problems in various fields for which there is no analytical approach. Since certain constants are believed to be fixed, even problems with analytical solutions have them. The analytical solution to a simplistic dilemma, on the other hand, teaches us a lot about the system's actions. On the other hand, if no analytical solution approach is available, numerical methods can be used to investigate problems rapidly. However, extreme caution must be exercised to ensure that a converged solution is achieved. This means we need to figure out if the step sizes are minimal enough to discover the solutions to the equations we are trying to understand. The numerical method FEM is an excellent tool to solve complicated geometrical shapes with a boundary and load condition that is difficult to describe with analytical expressions available. There are generally three approaches by which scientific problems/equations are solved: Analytical, Numerical, and Experimental. However, we cannot perform the experimental method every time because of cost and time constraints. The traditional approaches for solving problems are analytical methods. However, we cannot solve equations analytically due to limitations imposed by complex geometry, boundary conditions, and other factors.

Consequently, we have been pushing towards numerical methods for several years because they can produce almost reliable results in addition to analytical methods, and they can do so in a much shorter and easier period. The famous Navier-Stoke equation has never been solved analytically, but it can be solved quickly using Numerical Schemes. The disadvantages of traditional Numerical Schemes are they are challenging in representing irregular boundaries, not optimized for unstructured meshes, and momentum, energy, and mass are not conserved[10]. Lastly, they are not well suited for turbulent flow slow for significant problems, and they tend to be biased towards edges and one-dimensional physics.

To solve the drawbacks of traditional methods, we investigate the use of physics-informed neural networks (PINNs) as a solution approximation for PDEs in this paper. PINNs – neural networks trained to solve supervised learning tasks while respecting any given physics law described by general nonlinear PDEs Fig. 1 and Eq. (1) summarizes the PINNs[11]. We will talk about PyTorch as a Python implementation for PINNs. We assume these PINNs would affect real-world applications where reduced-order physics models are possible and widely used. The numerical efficiency of reduced-order models usually comes at the expense of physical consistency. We illustrate how PINNs can be used in reduced-order models to reduce epistemic (model-form) ambiguity for the advection equation. PINNs architectures can help close the distance between forecasts and observable results. Our method begins with an abstract formulation and then moves on to numerical integration methods before arriving at a neural network implementation.

$$\text{PINNs} = \text{Data} + \text{Neural Networks} + \text{Physical Laws} \qquad \text{Eq. (1)}$$

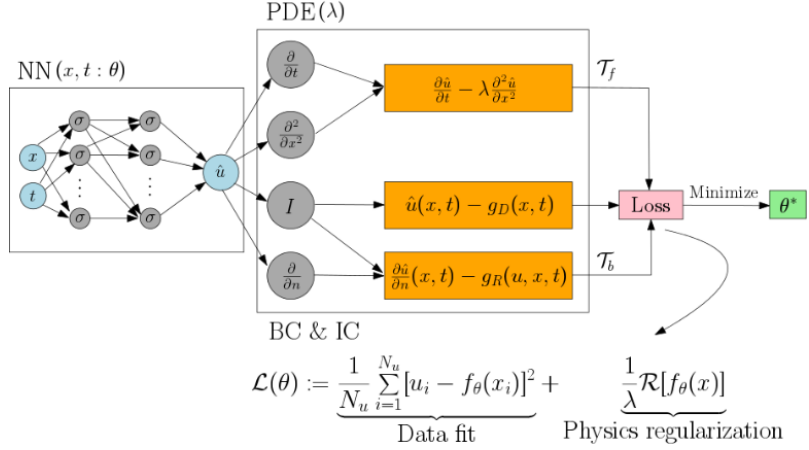

Fig.1. Schematic of a PINN for solving the diffusion equation ($\partial u/\partial t = \lambda\, \partial^2 u/\partial t^2$)

The rest of the paper is structured in the following way. Section 2 background of advection equation, section 3.1 specifies the implementation choices in terms of language and libraries, and public repositories (needed for replicating results). Section 3.2 presents the formulation and implementation for integrating the first-order ordinary differential equation with the simple Euler's forward method. Section 4 presents results, and section 5 closes conclusions and future work. Finally, Appendix summarizes concepts about neural networks used in this paper.

## 2. Background

Whether it is waves on a lake, sound waves, or an earthquake, everybody has seen traveling waves. Writing down a reasonable-looking finite difference approximation to a wave equation is not complicated. Most of the time is spent attempting to decide if the process succeeds. Since numerical wave propagation has its own set of complexities, we will start with one of the most basic mathematical equations for generating moving waves[12]. The advection equation is written as

$$\partial u/\partial t + a\, \partial u/\partial x = 0,\ for\ \begin{cases} -\infty < x < \infty \\ 0 < t \end{cases} \qquad \text{Eq. (2)}$$

The convenient feature of model Eq. (2) is that it has an analytical solution in Eq. (3):

$$u = u_0 f(x - at) \qquad \text{Eq. (3)}$$

Which represents a wave propagating with a constant velocity a with unchanged shape. When a>0, the wave propagating in the positive x-direction, whereas for *a<0*, the wave propagates in the negative x-direction. Equation (2) is significant in various applications should not be overlooked due to its mathematical simplicity. The momentum equation for gas motion, for example, is an inhomogeneous version of equation (2), that is dependent on u. Therefore, in the physical model of interstellar gas and its motion due to the solar wind, a nonlinear version of the advection equation occurs[13]. Another application is in analyzing traffic flow along a highway[14]. The density of cars along the road is represented by *u (x, t)*. Their speed is represented by a. Equation (2) may serve as a model-equation for a compressible fluid, e.g., if u denote pressure, it represents a pressure wave propagating. In a compliant pipe, such as a blood vessel, the advection equation is often used to model pressure or flow transmission. The applications of fractional advection-dispersion equations for anomalous solute transport in surface and subsurface water. The generalization of equation $\partial u/\partial t + \partial F/\partial x = 0$, where for the linear advection equation $F(u) = au$.

PDEs are equations that contain unknown multivariate functions and their partial derivatives in Eq. (4).

$$\emptyset_t + \nabla \cdot (u \cdot \emptyset) = \nabla \cdot (\Gamma \nabla \emptyset) \qquad \text{Eq. (4)}$$

Where $\emptyset$ is the dependent variable, $\emptyset_t = \partial \emptyset/\partial t$ is the derivative of $\emptyset_t$ concerning time, $t$, $\nabla = \left(\partial \emptyset/\partial t, \partial \emptyset/\partial t, \frac{\partial \emptyset}{\partial t}\right)$ is the nabla operator, $u = (u, v, w)$ is the velocity, and $\Gamma$ is the diffusivity. The independent variables are space, $x = (x, y, z)$, and time, $t$. Equation (4) is known as the convection-diffusion equation. Diffusion and convection are two mechanisms that explain how particles, electricity, and other physical quantities are transported within a physical structure. Concerning diffusion, $\nabla \cdot (\Gamma \nabla \emptyset)$, The concentration of a chemical can be assumed by $\emptyset$. As concentration is minimal in one region relative to the surrounding areas (e.g., a local minimum of concentration), the substance diffuses from the surrounding areas, increasing concentration. The net diffusion is proportional to the Laplacian (or second derivative) of concentration. if the diffusivity $\Gamma$ is a constant, $\Gamma \nabla^2 \emptyset$. On the other hand, concerning convection, $\nabla \cdot (u \cdot \emptyset)$, Assume the chemical is being carried down a canal, and we are calculating the concentration of the water per second. Someone spills a barrel of the chemical into the river upstream. The concentration will suddenly rise and fall as the field with the elevated chemicals passes. The problems presented in this study as examples will consist of finding the function $\emptyset(x, t)$ that, for a given geometry, original, and boundary conditions, satisfies the PDE. In a typical approach such as finite volume, we divide the statistical domain into small regions and consider the average volume size $\emptyset$ at a given moment Eq. (5).

$$\emptyset_i^n = \frac{1}{V_i} \int_{V_i} \emptyset(x, t^n) \qquad \text{Eq. (5)}$$

Where $V_i$ denotes the volume of the $i^{th}$ discretized element in the computational domain. The global solution can then be obtained by combining the different solutions for all volumes Eq. (6).

$$\emptyset^n = \sum_{V_i} \emptyset_i^n \qquad \text{Eq. (6)}$$

This function is piecewise constant and cannot be deduced, demonstrating one of the first drawbacks of conventional numerical methods. Many applications need derivable PDE solutions.

Specific exciting effects, such as heat fluxes and mass transfer, are computed with derivatives; a derivable solution would be more reliable than a derivatives approximation using piecewise functions. We discretize the various operators in equation one and use a time integration scheme to change the time. In its most basic form, the first-order Euler algorithm Eq. (7),

$$\emptyset_i^{n+1} = \emptyset_i^n - \frac{\Delta t}{V_i} \sum_{f_i} F_f A_f \qquad \text{Eq. (7)}$$

where $\Delta t$ is the time step, $F_f$ is the flux $\emptyset$ across the face f of the volume $V_i$ and $A_f$ is the area of the face. Additional computational schemes to calculate these unknown values are used to calculate the fluxes at the faces. Some popular choices are central difference or upwind methods. Two more aspects are required to complete the problem: the original and boundary conditions. To begin, keep in mind that an initial state is simply a time-dimensional boundary condition. This is important because initial and boundary conditions are typically considered independently, but they will be treated similarly in our approach. The initial condition sets the value for $\emptyset(x, t = 0)$ and functions as the first values to start the computation in the previously described time-marching algorithm. Boundary conditions set the value for $\emptyset(x \in D, t > 0)$ where $D$ is the total number of points on the domain's boundaries. We must specify a specific set of rules to change these points, not to see values outside the scope. There are many boundary conditions, but the ones used in our examples are as follows:

- Periodic: The domain folds itself to bind boundaries when periodic conditions are assumed.
- Dirichlet: For this type of boundary condition, we will fix $\emptyset$ it at boundaries.
- Newmann: For this type of boundary condition, we will fix $\nabla\emptyset$ at boundaries.

For the treatment of walls, inflows, and outflows, specific boundaries may be needed. In the temporal dimension, an initial state is a Dirichlet boundary condition. The approach used for other conventional approaches such as finite difference or finite elements differs slightly. The basic principle remains the same: discretize the computational domain into small regions where the solution is assumed and bring them back together to retrieve the global solution. Therefore, there are no derivable piecewise alternatives. Furthermore, since we use time-marching algorithms, new computations are needed once the free-parameters, current, or boundary conditions are changed.

## 3 Methods:

### 3.1 Implementation

This part will teach you how to solve equation (2) using a neural network. Python is being used to solve the PDEs. Python, a programming language, has grown in popularity in recent years. For the following reasons: Since there is no need for an explicit declaration of variables or a separate compilation period, it is quick and simple to code and use for small "prototyping" tasks. It is available for free on most computing systems and has a vast repository of packages covering a wide range of applications. Python also has features that make developing and documenting massive, well-structured program structures easier. Python is not appropriate for running extended computational computations since it is an interpreted language. However, calls to precompiled library routines are often used in the code for such computations. Many of these routines are available directly from Python using the PyTorch[15], NumPy[16], and SciPy[17] packages. Most operating systems (OS), including Linux, OSX, and MSWindows, have these kits available for free.

The aim is to achieve a trained multilayer perceptron (MLP) that can give the output *φ(x, t),* where *x* and *t* are set as inputs that satisfy equation 1. The basic principle is that a forward transfer on the network using the independent variables as PINN inputs gives one the value of the dependent variables evaluated. Since PINNs are derivable, we can compute the dependent variables (outputs). The dependent variables (inputs) to calculate the different derivatives that appear in the original PDEs. We create a loss function that fits the PDEs with this derivative, and we use it throughout the training phase. We will assume that our PINNs are a solution to the PDEs if the loss function approaches a near-zero value. The teaching is done in an unsupervised setting. There are continuous and derivable solutions that can be applied throughout the whole domain. PINNs also have the advantage of allowing one to use the PDEs-free parameters in the solution. Therefore, a solution trained for various values of these parameters will generalize to different conditions rather than a single scenario, eliminating the need for new calculations any time a parameter is modified. This property is of particular interest in optimization studies. In more depth, we describe a collection of points within our domain in the same way as standard approaches would. These points are divided into two categories: preparation and validation after training. We also differentiate between internal and external points. This will be dealt with in compliance with the boundary conditions that have been established. Then, we define the MLP architecture: several layers and a number of hidden units in each layer. The number of inputs would be the same as the number of independent variables in the PDEs with any free parameters we want to add. The number of outputs would be the same as the number of unknowns that need to be solved. Table 1 summarizes the PINNs algorithm. Once we have training data and the PINNs defined, follows the steps:

- We compute the network's outputs at all points, $\emptyset(x, t)$ and the derivatives concerning the inputs: $\emptyset_t, \emptyset_x, \emptyset_{xx}$.
- We use a loss function that fits our PDE for internal points. This is the role we'd like to improve: $\emptyset_t + \nabla \cdot (u \cdot \emptyset) - \nabla \cdot (\Gamma \nabla \emptyset) = 0$
- We can create an MSE loss function to fulfill the given condition for boundary points since we fix values.
- Update the parameters of the PINNs for each loss function.

Table 1: PINNs Algorithm.

| **PINNs Algorithm** |
|---|
| **Input:** dynamics parameters $\emptyset$, start time $t_0$ , stop time $t_1$, final state z $(t_1)$.<br>• Specify the two training sets $T_f$ and $T_b$ for the equation and boundary/intial conditions.<br>• Define dynamics on augmented state.<br>• Specify a loss function for PDE equation and boundary condition.<br>• Train the PINNs to find the best parameters $\emptyset$ by minimizing the loss function. |

**3.2 Solving one-dimensional advection equation using PINNs.**

We aim to create a qualified multilayer perceptron (MLP) that can output *(x, t)* when given *x* and *t* as inputs and satisfy equation (2). Consider the one-dimensional advection Eq. (8), which is a simplification of equation (1) for a 1D,

$$\emptyset_t + u\emptyset_x = 0 \qquad \text{Eq. (8)}$$

Where $\emptyset(x,t)$ is the unknown function, $x$ and $t$ are the independent variables, $u$ is a constant parameter and $\emptyset_t$ and $\emptyset_x$ are the derivatives of $\emptyset$ concerning $t$ and $x$, respectively. This PDE has an analytical solution, which is $\emptyset(x,t) = \emptyset(x, x - ut)$. We may assume that the initial condition was physical $\emptyset(x, t = 0)$ moves in x at speed u. In the case that $\emptyset(x, t = 0) = exp\left(-\frac{1}{2}\left(\frac{x-at}{0.4}\right)^2\right)/L$ the solution is $\emptyset(x,t) = exp\left(-\frac{1}{2}\left(\frac{x-at}{0.4}\right)^2\right)/L$ as illustrated in Fig. 3. To solve the equation, we define a set of points for training.

$$\text{Discretized equation: } u_i^{n+1} = u_i^n - a\frac{\Delta t}{\Delta x}(u_i^n - u_{i-1}^n) \qquad \text{Eq. (9)}$$

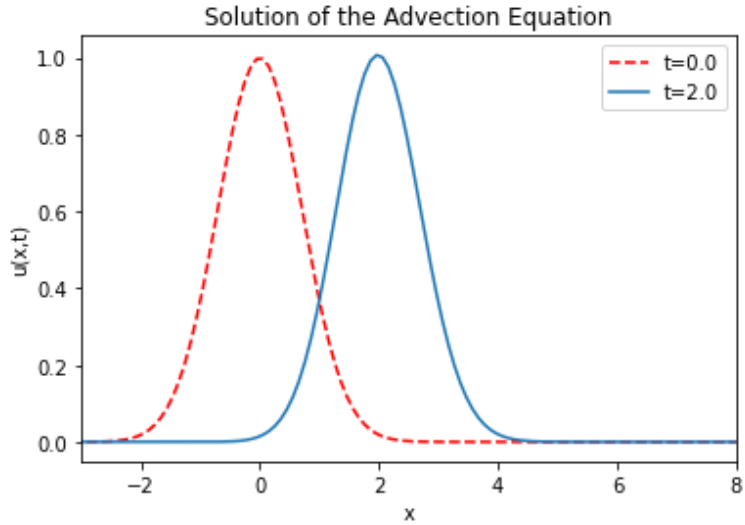

Fig. 3. Example of the solution to the 1D advection equation with the initial condition $exp\left(-\frac{1}{2}\left(\frac{x-at}{0.4}\right)^2\right)$

Defining a $\Delta x$ and $\Delta t$ allows us to build a uniform grid of points in the entire domain used in the discretized Eq. (9). We describe internal and boundary points, each of which would have a separate loss function associated with it. When t = 0, the initial condition will use a Mean Square Error (MSE) loss function, which will equate the known initial condition with the PINNs output. For the spatial boundary condition, we use a periodic condition that compares the solutions using an MSE loss function at *x = 0* and *x = L* for any t and forces them to be equal. As seen in Fig. 4, we characterize our solution as an MLP with two inputs (number of independent variables), hidden layers, and one output. Fig. 4 shows a general scheme's flowchart for converting a differential equation into the PINNs structure.

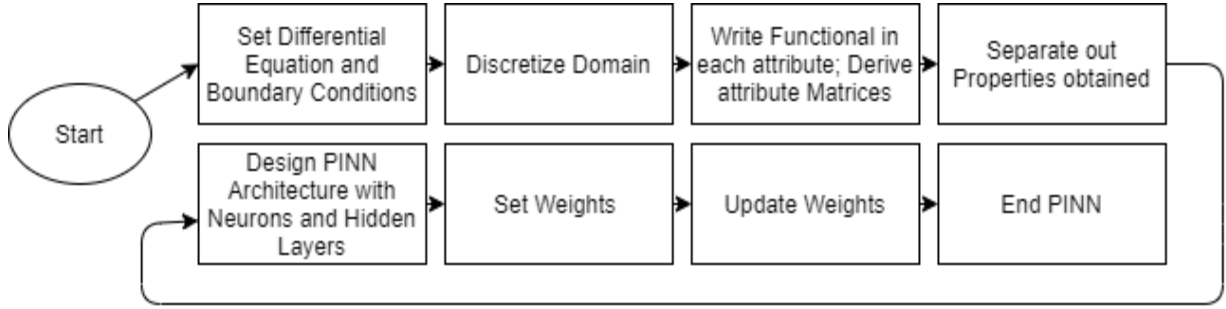

Fig. 4. Flowchart for designing the PINNs for a general PDE

The training steps as follows:
- Compute the network outputs for the internal points: $\emptyset(0 < x < L; t > 0)$.
- Compute the gradients of the outputs concerning the inputs: $\emptyset_x, \emptyset_t$.
- Create the internal point loss function: L1 =MSE $(\emptyset_x + \emptyset_t)$
- Calculate boundary state outputs: $\emptyset(0 < x < L; t = 0)$, $\emptyset(x = 0; t)$ and $\emptyset(x = L, t)$.
- Establish a failure function to account for boundary conditions: L2 =MSE $(\emptyset(0 < x < L; t = 0) - exp\left(-\frac{1}{2}\left(\frac{x-at}{0.4}\right)^2\right)/L$, L3 = MSE $(\emptyset(x = 0; t) - (\emptyset(x = L; t))$
- Update the PINNs parameters for the other losses.

MLP with two inputs, $x$ and $t$, one output $\emptyset(x,t)$, sigmoid activation function, and five hidden layers with 32 neurons shown in Fig. 5. Moreover, we here assume that $t$ is a function of $\emptyset(x,t)$, the first derivative of $\emptyset(x,t)$ concerning the set of inputs and physical parameters, $\theta$. The training points are randomly uniformly drawn from their corresponding domains. $\{x^{n,i}, u^{n,i}\}$ corresponding to data at $t^n$ and Euler scheme equation (7) now allows us to infer the latent solution $u(t, x)$ in a sequential method. Starting from initial data $\{x^{n,i}, u^{n,i}\}$ at time $t^n$ and data at the domain boundaries x = 0 and x = 2 shown in Fig. 5, we can use the loss function to train the networks and predict the solution at the time $t^{n+1}$. A Euler time-stepping scheme would then use this prediction as initial data for the next step and proceed to train again and predict. We use an optimization algorithm to perform well in applying the PINNs method for parameter estimation. The Adam optimizer with a learning rate of 0.001 is used until the optimization problem's solution converges with the prescribed tolerance. We set the maximum number of Adam's epoch to be 10,000 and use the mini-batch training; the PINNs weights are randomly initialized using the Xavier scheme[18].

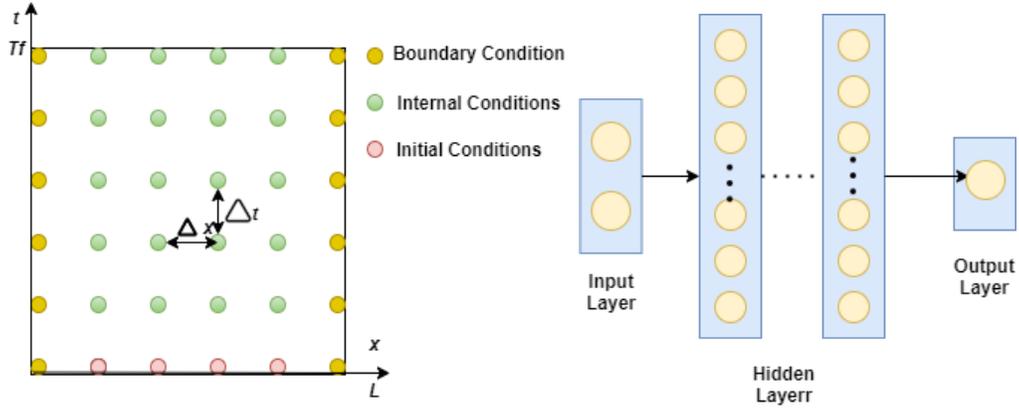

Fig. 5. Two distributions of training points of PINNs for one-dimensional advection equation (Left). PINNs Multilayer perceptrons (Right)

These steps are usually small in the classical numerical analysis due to stability constraints for explicit schemes or computational complexity constraints for implicit formulations. If the number of Euler scheme stages increases, these restrictions become more serious. Thousands to millions of such measures are needed for most realistic interest problems before the solution is resolved up to a desired final period. In comparison to conventional approaches, we can use an implicit Euler scheme with an infinitely large number of stages at no added expense. This helps one to overcome the entire Spatio-temporal solution in a single step while ensuring consistency and high predictive precision. Table 2 shows the parameters and constants used implementation of PINNs and different hyperbolic schemes.

Table 2: Constants and parameters

| Constants and parameters | Values |
| --- | --- |
| a (Wave Speed) | 1.0 |
| $t_{min}$, $t_{max}$ (start and stop time of simulation) | 0.0,2.0 |
| $N_x$ (Number of spatial points) | 80 |
| C (Courant number, need C<=1 for stability) | 0.9 |

## 4. Results

A one-dimensional advection equation is solved to understand the main process of solving a PDE with a PINNs. Compared PINNs with a traditional FTBS [19], Lax Wendroff[20], and Lax Friedrich[21] all time-integrated with an Euler scheme. Our trained PINNs, unlike the FVM solution, are continuous and derivable over the entire domain. Since physical effects do not limit PINNs, our experiments demonstrate that a much larger mesh will provide better results in this specific case. In comparison to conventional methods, we find that PINNs are capable of minimizing error. Our findings are summarized in table 3 below.

Table 3: MSE reported. MSE between the predicted $\emptyset(x,t)$ and the exact solution *u(x, y)* for the different methods and PINNs.

| Methods | MSE |
|---|---|
| FTBS | 3.87e-03 |
| Lax Wendroff | 4.32e-03 |
| Lax Fredrich | 4.78e-03 |
| PINNs | 1.65e-03 |

The computing time needed to integrate the problem in a small grid of points is a few seconds. Still, if the domain's discretization contains about 10000 points for each variable, the computing time of the 10000 Euler method is not similar to that of the PINNs training. However, the computing time requirement of the PINNs method is not comparable with traditional numerical methods. Any numerical method gives solution points over a selected grid. The computing time, in this case, increases as the number of grid points increases. So, computing the solution for an infinite number of points will need infinite time. On the other hand, the PINNs find a continuous solution over the real domain in a finite time.

The PINNs were tested with points that did not participate in the training. Fig. 5 shows the MSE of the trained network as a function of the number of testing points. It can be observed that the averaged error is stable for an increasing number of testing points, which indicates that the solution found is a good approximation of the real solution not only for the training points but also over the whole domain of the problem.

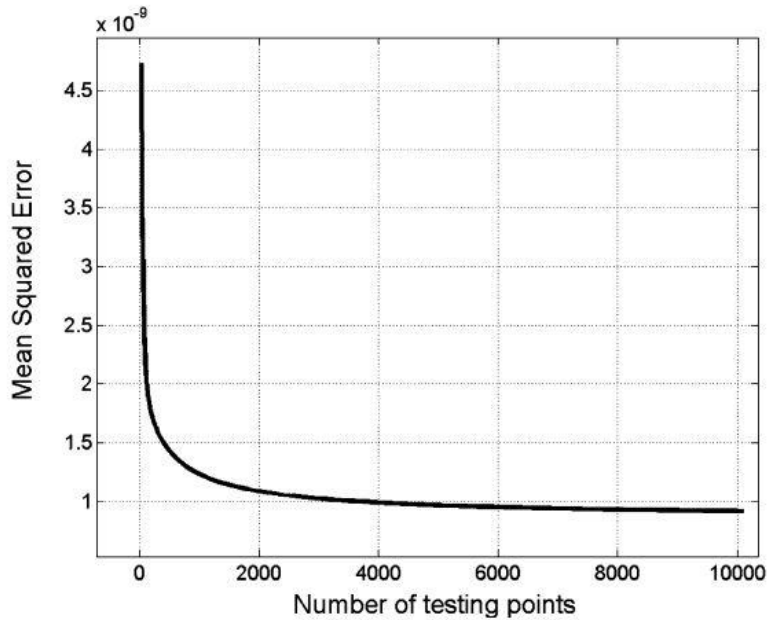

Fig. 5. MSE of the trained PINNs vs the number of testing points into the integration domain.

Fig. 6 shows the errors in the resulting PINNs solution $\emptyset(x, t)$ for the reference one-dimensional advection equation *u(x, t)* field with the maximum point errors provided in Table 4 for the prediction in *0 < t < 2*. The FTBS, Lax Wendroff, and Lax Friedrich approximation produced a less accurate solution and have more intense oscillation than PINNs. As can be seen, training the PINNs requires optimization concerning many loss functions (as many as PDEs and different boundary conditions), challenging complex problems. We see that PINNs more sharply resolve discontinuities at the cost of adding oscillations, which leads to a net reduction in error. We can use this result to hypothesize that PINNs will further improve the solution when the discontinuity is more significant. We verify our hypothesis by plotting the error ratio against discontinuity width in *0 < t < 2* and demonstrating that more significant discontinuities lead to lower error ratios. In the solutions obtained with Lax Friedrich, FTBS, and Lax Wendroff, an anomalous oscillation is present when the grid spacing is too large to follow the quantity's variations advected closely. Lax Friedrich, FTBS, and Lax Wendroff results converge faster and produce more significant predictive errors than PINNs at almost all training data points. The accuracy of the predictions Lax Friedrich, FTBS, and Lax Wendroff is very good at the initial conditions *t=0,* and the boundary conditions are controlled or well known, but they struggle at *t=2*. MSE indicates that the PINNs loss can produce an excellent approximation to the solution (*0 < t < 2*) without explicit regularization.

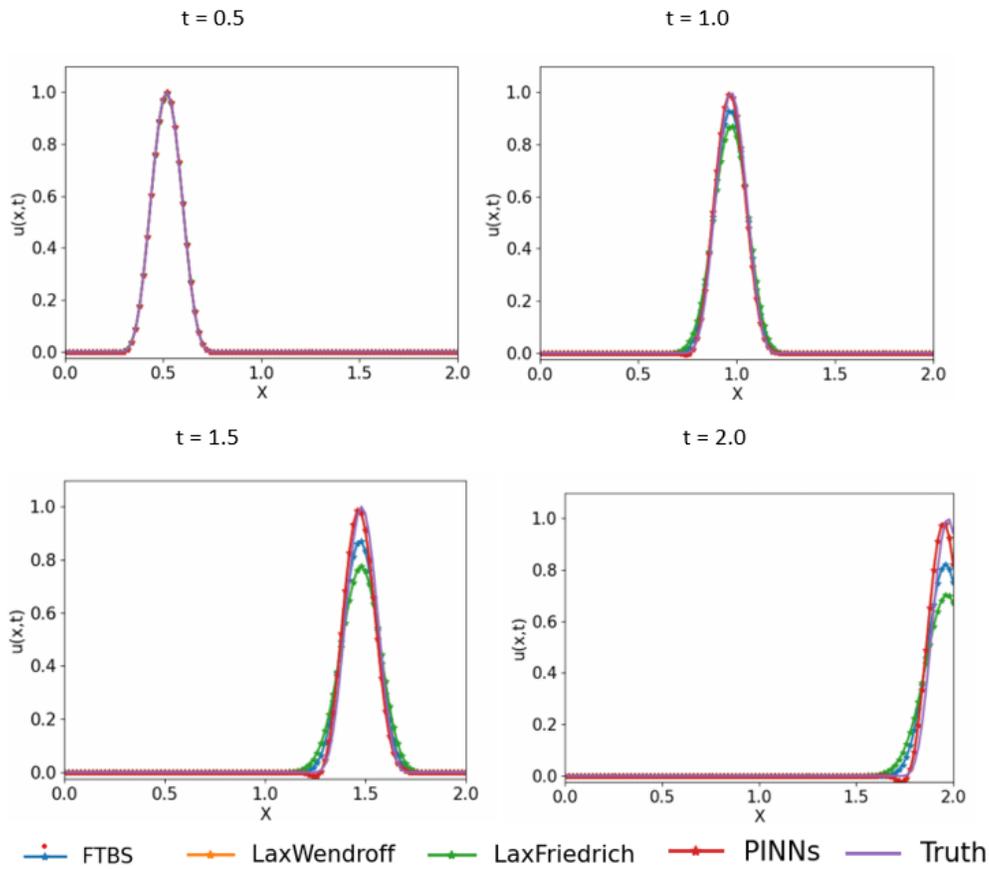

Fig. 6: A range of hyperbolic schemes are implemented 1D advection equation with the initial condition $exp\left(-\frac{1}{2}\left(\frac{x-at}{0.4}\right)^2\right)$. The colored lines with a marker "*" denote different hyperbolic schemes. The solid black line denotes the exact solution.

Table 4: 1D advection equation: Relative final prediction error measure in the L2 norm for different hyperbolic schemes. Here, the time-step size to t = 0.5.

| Model | t = 0 min | t = 0.5 | t = 1 | t = 1.5 | t = 2 |
| --- | --- | --- | --- | --- | --- |
| FTBS | 2.33e-04 | 5.29e-04 | 5.46e-03 | 6.01e-03 | 7.12e-03 |
| Lax Wendroff | 2.03e-04 | 4.90e-04 | 5.26e-03 | 6.71e-03 | 8.92e-03 |
| Lax Friedrich | 2.53e-04 | 5.80e-04 | 6.56e-03 | 7.21e-03 | 9.32e-03 |
| PINNs | 1.03e-04 | 1.10e-7 | 2.26e-03 | 2.45e-03 | 3.42e-03 |

## 5. Conclusions:

By comparing numerical solutions of advection equation with known analytic solutions and PINNs have been demonstrated. The FTBS, Lax Wendroff, and Lax Friedrich introduced a pseudo diffusion effect that led to a large error in *t=2* and would make accurate turbulent diffusion modeling impossible. In those cases where the value of the property being advected varied rapidly in the space of a few grid intervals, variations which seem to be typical in many models, the Lax Friedrich, FTBS, and Lax Wendroff introduced an anomalous oscillation into the distribution of the quality being advected at *t=2*. The oscillations grew steadily with time and could easily introduce instability into a numerical model. Of the methods investigated, Only the PINNs approximation moved the advected field correctly at *0 < t < 2*. Traditional numerical methods to find a consistent solution with more straightforward and faster methods led to inaccurate results. Only the PINNs method produces an accurate and consistent approximation with the PDEs. Traditional numerical methods also require a large amount of computer core storage. PINNs approach can transform the physics simulation area by allowing real-time physics simulation and geometry optimization without costly and time-consuming simulations on large supercomputers.

multifidelity neural network (MFNN) framework," *J. Rheol.*, vol. 65, no. 2, pp. 179–198, Feb. 2021, doi: 10.1122/8.0000138.

[5] J. Xu, Q. Tao, Z. Li, X. Xi, J. A. K. Suykens, and S. Wang, "Efficient hinging hyperplanes neural network and its application in nonlinear system identification," *ArXiv190506518 Cs*, Nov. 2019, Accessed: Mar. 14, 2021. [Online]. Available: http://arxiv.org/abs/1905.06518.

[6] A. Albalowi and A. Alhamed, *Big data and learning analytics in higher education: Demystifying variety, acquisition, storage, NLP and analytics*. 2017.

[7] K. Denecke, "Extracting Medical Concepts from Medical Social Media with Clinical NLP Tools : A Qualitative Study," 2014. /paper/Extracting-Medical-Concepts-from-Medical-Social-NLP-Denecke/8ed7609f41aa772a93c4a46f5ae7a9266559376e (accessed Mar. 14, 2021).

[8] E. Strubell, A. Ganesh, and A. McCallum, "Energy and Policy Considerations for Deep Learning in NLP," *ArXiv190602243 Cs*, Jun. 2019, Accessed: Mar. 14, 2021. [Online]. Available: http://arxiv.org/abs/1906.02243.

[9] S. R. Vadyala, S. N. Betgeri, E. A. Sherer, and A. Amritphale, "Prediction of the Number of COVID-19 Confirmed Cases Based on K-Means-LSTM," *ArXiv200614752 Phys. Q-Bio*, Jun. 2020, Accessed: Mar. 14, 2021. [Online]. Available: http://arxiv.org/abs/2006.14752.

[10] C. Loredo, D. Banks, and N. Roqueñí, "Evaluation of analytical models for heat transfer in mine tunnels," *Geothermics*, vol. 69, pp. 153–164, Sep. 2017, doi: 10.1016/j.geothermics.2017.06.001.

[11] M. Raissi, P. Perdikaris, and G. E. Karniadakis, "Physics-informed neural networks: A deep learning framework for solving forward and inverse problems involving nonlinear partial differential equations," *J. Comput. Phys.*, vol. 378, pp. 686–707, Feb. 2019, doi: 10.1016/j.jcp.2018.10.045.

[12] D. Dutykh and F. Dias, "Water waves generated by a moving bottom," in *Tsunami and Nonlinear Waves*, A. Kundu, Ed. Berlin, Heidelberg: Springer, 2007, pp. 65–95.

[13] R. Bruno and V. Carbone, "The Solar Wind as a Turbulence Laboratory," *Living Rev. Sol. Phys.*, vol. 10, no. 1, p. 2, May 2013, doi: 10.12942/lrsp-2013-2.

[14] L. Romero and F. G. Benitez, "OUTLINE OF DIFFUSION-ADVECTION IN TRAFFIC FLOW MODELLING," p. 17.

[15] E. Stevens, L. Antiga, T. Viehmann, and S. Chintala, *Deep learning with PyTorch*. 2020.

[16] T. Oliphant, *Guide to NumPy*. 2006.

[17] "Learning SciPy for Numerical and Scientific Computing," *Packt*. https://www.packtpub.com/product/learning-scipy-for-numerical-and-scientific-computing/9781782161622 (accessed Mar. 14, 2021).

[18] S. K. Kumar, "On weight initialization in deep neural networks," *ArXiv170408863 Cs*, May 2017, Accessed: Mar. 14, 2021. [Online]. Available: http://arxiv.org/abs/1704.08863.

[19] W. Long, J. Kirby, and Z. Shao, "A numerical scheme for morphological bed level calculations," *Coast. Eng.*, vol. 55, pp. 167–180, Feb. 2008, doi: 10.1016/j.coastaleng.2007.09.009.

**Appendix. PINNs method**

PINNs is a simple multilayer feedforward neural network with depth $D$ that contains an input layer, $D - 1$ hidden layers, and output layer. Without loss of generality, we assume that there are $N_d$ neurons in the $d^{th}$ hidden layer. Then, the $d^{th}$ hidden layer receives the post-activation output $x^{d-1} \in R^{N_{d-1}}$ of the previous layers as its input, and the specific affine transformation is of the form shown in Eq. (A.1):

$$\rho_d(X^{d-1}) \triangleq W^d x^{d-1} + b^d \qquad \text{Eq. (A.1)}$$

where the network weight $W^d \in R^{N_{d-1}}$ and the bias term $b^d \in R^{N_d}$ be learned are initialized using some special strategies, such as Xavier initialization or initialization. The nonlinear activation function σ(·) is applied component-wise to the affine output $\rho_d$ of the present layer. For specific regression issues, this nonlinear activation is not used in the output layer, or we may assume that the identity activation is used in the output layer. As a result, the neural network can be denoted as $q(x, \Theta) = (\rho_d \circ \sigma \circ \rho_{d-1} \circ \cdots \circ \sigma \circ \rho_1)(x)$. Where the operator " ∘ " is the composition operator, $\Theta = \{W^d, b^d\}^D \in$ P represents the learnable parameters to be optimized later in the network, and P is the parameter space, q, and $x^0$ = x is the output and input of the network, respectively. The residual networks $f_u(x, t) := u_t + N_u(u, u_x, u_{xx}, u_{xxx}, \cdots)$, and $f_v(x, t) := v_t + N_v(v, v_x, v_{xx}, v_{xxx}, \cdots)$. Then the solution $q(x, t)$ will be trained to satisfy these two physical constraint conditions. Which play a vital role in regularization and have been embedded into the mean-squared objective function, that is, the loss function $loss_\Theta = loss_u + loss_v + loss_{f_u} + loss_{f_v}$

**Appendix. Multilayer perceptrons (MLP) shown in Fig. A.1**

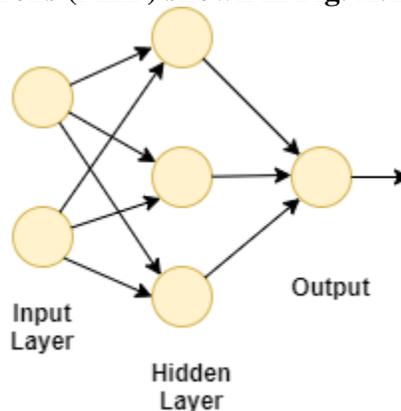

Fig. A.1 illustrates the popular MLP.

Each layer can have one or more perceptrons (nodes in the graph). A perceptron applies a linear combination to the input variables followed by an activation function.

$$v = f(z) \text{ and } z = W^T u + b$$

where $v$ is the perceptron output, $u$ is the inputs; $w$ and $b$ are the perceptron hyperparameters, and $f(.)$ is the activation function. Throughout this paper, we used the hyperbolic tangent (tanh), sigmoid, and the exponential linear unit (elu) activation functions (although others could also be used, such as the rectified exponential linear unit):

$$\tanh(z) = \frac{e^z - e^{-z}}{e^z + e^{-z}}, \text{ sigmoid}(z) = \frac{1}{1+e^{-z}} \text{ and } \text{elu}(z) = \begin{cases} z & \text{when } z > 0 \\ e^z - 1 \end{cases}$$